\documentclass[9pt,shortpaper,twoside,web]{ieeecolor}
\usepackage{generic}
\usepackage{cite}
\usepackage{amsmath,amssymb,amsfonts}
\usepackage{algorithmic}
\usepackage{graphicx}
\usepackage{textcomp}
\usepackage{url}
\let\proof\relax
\let\endproof\relax
\usepackage{amsthm}
\usepackage{cancel}
\usepackage{xcolor}

\def\BibTeX{{\rm B\kern-.05em{\sc i\kern-.025em b}\kern-.08em
    T\kern-.1667em\lower.7ex\hbox{E}\kern-.125emX}}
\theoremstyle{plain}% Theorem-like structures provided by amsthm.sty
\newtheorem{theorem}{Theorem}
\newtheorem{lemma}{Lemma}

\newtheorem{proposition}{Proposition}
\theoremstyle{proposition}
\theoremstyle{assumption}
\newtheorem{definition}{Definition}

\newtheorem{remark}{Remark}
\theoremstyle{goal}

\theoremstyle{assumption}
\newtheorem{assumption}{Assumption}

\begin{document}
\title{Instrumental Variables based DREM for Online Asymptotic Identification of Perturbed Linear Systems}
\author{Anton Glushchenko, \IEEEmembership{Member, IEEE}, Konstantin Lastochkin
\thanks{A. I. Glushchenko is with V.A. Trapeznikov Institute of Control Sciences RAS, Moscow, Russia (phone: +79102266946; e-mail: aiglush@ipu.ru).}
\thanks{K. A. Lastochkin is with V.A. Trapeznikov Institute of Control Sciences RAS, Moscow, Russia (e-mail: lastconst@ipu.ru).}}

\maketitle

\begin{abstract}
Existing online continuous-time parameter estimation laws provide exact (asymptotic/exponential or finite/fixed time) identification of dynamical linear/nonlinear systems parameters only if the external perturbations are
equaled to zero or {independent with the regressor of the system. However, in real systems the disturbances are almost always non-vanishing and dependent with the regressor.} In the presence of perturbations with such properties the above-mentioned identification approaches ensure only boundedness of a parameter estimation error. The main goal of this study is to close this gap and develop a novel online continuous-time parameter estimator, which guarantees exact asymptotic identification of unknown parameters of linear systems in the presence of unknown but bounded perturbations and has {relaxed convergence conditions}. To achieve the aforementioned goal, it is proposed to augment the deeply investigated Dynamic Regressor Extension and Mixing (DREM) procedure with the novel Instrumental Variables (IV) based extension scheme with averaging. Such an approach allows one to obtain a set of scalar regression equations with asymptotically vanishing perturbation if the initial disturbance that affects the plant is bounded and {independent not with the system regressor, but with the instrumental variable.} It is rigorously proved that a gradient estimation law designed on the basis of such scalar regressions ensures online unbiased asymptotic identification of the parameters of the perturbed linear systems if {some weak independence and excitation assumptions are met}. Theoretical results are illustrated and supported with adequate numerical simulations.
\end{abstract}


\section{Introduction}
\label{sec:introduction}
The aim of the adaptive control methods is a real-time control of dynamical systems with unknown parameters \cite{b1, b2, b3}. As far as the methodology of this theory is concerned, a control law is designed in two stages. At the first one, an ideal control law is chosen that ensures the achievement of the control goal under the assumption that the system parameters are known. At the second stage, in accordance with the \emph{certainty equivalence} principle, the unknown parameters of the chosen control law are substituted with their dynamic estimates obtained with the help of adaptive laws, which are designed using direct or indirect approaches. According to the direct method, the adaptive law is designed to adjust the controller parameters directly. In accordance with the indirect approach, first, the system parameters are identified, and then they are recalculated into the controller parameters using algebraic equations. To design the adaptative law, in both cases the second Lyapunov method \cite{b2} and parametric identification algorithms (the gradient law \cite{b4}, various variants of the recursive least squares method \cite{b5}) are used.

Long standing question in the sense of the above-mentioned identification and adaptive control problems is to define conditions to guarantee robustness of a parameter estimator or/and overall closed-loop adaptive control system to external bounded perturbations, unknown dynamics and system parameter variations \cite{b6, b7, b8, b9}. It is well known \cite{b6} that the basic adaptive and identification laws ensure uniform ultimate boundedness (UUB) of the tracking and parameter estimation errors in case when the persistence of excitation (PE) requirement with sufficiently large level of excitation is met. However, these results are impractical due to two following main problems.

\begin{enumerate}
\item [\textbf{ MP1.}]  Considering linear systems {that satisfy certain input stationarity assumptions}, the PE condition is met if and only if the system input includes $n$ not equal to each other frequencies, which is a restrictive requirement for the practical scenarios.
\item [\textbf{ MP2.}] Only UUB of the tracking and parameter estimation errors is ensured in the presence of external perturbations even if the PE condition is satisfied.
\end{enumerate}

To solve the first problem, the adaptive control community has proposed many robust modifications ($\sigma$, $e$, projection operator, dead zone etc.), which ensure required UUB \cite{b1, b2, b6, b7, b8, b9} even if the PE condition is not met. Moreover, to close \textbf{MP1} a great effort has recently been made \cite{b4, b10} to relax the PE condition to exponential stability requirement, which is sufficient \cite[p.~327]{b3} to ensure robustness (UUB) of the closed-loop adaptive control system in the presence of external perturbations. However, all approaches under consideration do not ensure convergence of the parameter estimation and/or tracking errors to zero even if an arbitrary small time-dependent non-vanishing perturbation affects the system, but this is \emph{a vita} necessary to achieve acceptable control quality and identification accuracy in practice (\textbf{MP2}). Considering the adaptive control framework, to cope with the external perturbations, a methodology \cite{b3, b11} has been developed that uses the internal model principle to parameterize a perturbation in the form of a linear regression equation with respect to some \emph{new} unknown parameters with higher dimension. However, such approach has three main drawbacks: \emph{i}) the overall closed-loop system or/and system parameterization are complicated due to \emph{overparametrization}, \emph{ii}) the internal models can be used for description of relatively small class of perturbations (constant, , exponentially decaying signals and their combinations), and \emph{iii}) \emph{a priori} knowledge of the disturbance type is required to apply the internal model principle.

Generally, the origin of sensitivity of the adaptive control systems to unknown but bounded time-dependent perturbations is rooted in the properties of the adaptive and estimation laws, which are applied to solve the adaptive control and identification problems. Particularly, to the best of authors’ knowledge, considering the continuous time adaptive control literature \cite{b1, b2, b3}, {only pure least-squares (P-LS) estimation law ensures exact asymptotic identification of the unknown parameters in the presence of perturbations, which are independent with the system regressor \cite{Fr}. However, the above-mentioned approach is hard to tune, provides only asymptotic rate of convergence even in the petrurbation-free case and ensures parameter convergence only if the restrictive PE condition is met and the system regressor is independent with perturbations.} Recently proposed robust identifiers with finite/fixed convergence time and relaxed excitation requirement \cite{b12, b13} also provide only boundedness of the parametric error in the presence of an external disturbance. So, the main topic of this study is to close this gap and propose continuous-time online estimation law for linear time-invariant plants, which in comparison with P-LS \cite{Fr} and other existing estimation laws \cite{b1, b2, b3, b4, b5, b6, b7, b8, b9, b10, b11, b12, b13, b14} has the following novel properties:
\begin{enumerate}
\item[\textbf{P1)}] in the perturbed case it ensures \emph{online} exact asymptotic estimation of the unknown parameters of linear systems even if the perturbation and regressor of the system are dependent,
\item[\textbf{P2)}] in perturbation-free case it provides the exponential rate of convergence,
\item[\textbf{P3)}] both in perturbed and perturbation free cases it has weaker condition of asymptotic convergence in comparison with PE.
\end{enumerate}

It is necessary to remember that, if the law with such property is obtained, then in the future it could potentially be \emph{mutatis mutandis} modified and applied to solve any kind of adaptive and identification problems for continuous-time not necessarily linear systems.

Now we give some small description of the mechanisms and approaches that will be applied in the paper to achieve the stated goal. First of all, using the state variables filters that are well-known in the adaptive control literature \cite{b1, b2, b3}, a linear time-invariant system affected by a disturbance is parameterized in the form of a linear regression equation (LRE) with unknown but bounded perturbation \cite[p.~49]{b1}. Then, the obtained LRE is extended with the help of a novel extension scheme, which is obtained via a combination of: \emph{i}) the sliding window extension scheme \cite[Lemma 1]{b14}, \emph{ii}) the instrumental variables method \cite{b15, b16, b17, b18, b19} and \emph{iii}) the filtering with averaging \cite{b20}. Such novel extension scheme ensures that the new perturbations in new regresion equation converge asymptotically to zero with the rate ${\textstyle{1 \over {t + {F_0}}}}{\rm{,\;}}{F_0} > 0$, {{\it e.g.}} if the system disturbance and control input do not include signals with common frequencies. The next ingredient of the proposed estimation design procedure is the mixing step \cite{b4} from the DREM method, which transforms the extended linear regression equation into a set of separate scalar regression equations and {invokes the relaxation of condition for convergence}. Being applied to the obtained set of scalar regression equations, the gradient-descent-based estimator ensures required properties \textbf{P1}-\textbf{P3}.

The rest of the study is organized as follows. A rigorous problem formulation is shown in Section {II}. The essence of instrumental variable approach is discussed in Section {III}. In Section {IV} the main result is elucidated. The simulation results are listed in Section {V.} {The main definitions of the harmonic analysis and proofs of the main results of the study are presented in Supplementary Material \cite{Supp}.}

\textbf{Notation and Definitions.} The below-given definitions and notation are used to present the main result of this study.
\begin{definition}\label{definition1} The regressor $ \varphi \left( t \right) \in {\mathbb{R}^n}$ is persistently exciting $\left( {\varphi \in {\rm{PE}}} \right)$ if $\exists T > 0$ and {$\overline{\alpha}\ge\underline{\alpha} > 0$} such that $\forall t \ge {t_0} \ge 0$ the following inequality holds:
\begin{gather*}
{\overline{\alpha}{I_n}\ge}\int\limits_t^{t + T} { \varphi \left( \tau  \right){{ \varphi }^{\top}}\left( \tau  \right)d} \tau {\ge \underline{\alpha}} {I_n}.
\end{gather*}
\end{definition}

\begin{definition}\label{definition2}
Signals $f\left( t \right) \in \mathbb{R}$ and $g\left( t \right) \in \mathbb{R}$ are called independent if {for all $t\ge t_0$} it holds that:
\begin{gather*}
\left| {\int\limits_{{t_0}}^t {f\left( s \right)g\left( s \right)ds} } \right| < \infty .
\end{gather*}

Otherwise, $f\left( t \right) \in \mathbb{R}$ and $g\left( t \right) \in \mathbb{R}$ are called dependent.
\end{definition}

Further the following notation is used: $\left| . \right|$ is the absolute value, $\left\| . \right\|$ is the suitable norm of $(.)$, ${I_{n \times n}}=I_{n}$ is an identity $n \times n$ matrix, ${0_{n \times n}}$ is a zero $n \times n$ matrix, $0_{n}$ stands for a zero vector of length $n$, ${\rm{det}}\{.\}$ stands for a matrix determinant, ${\rm{adj}}\{.\}$ represents an adjoint matrix. {We say that $f\in L_{q}$ if 
$\sqrt[q]{{\int\limits_{{t_0}}^t {{f^q}\left( s \right)ds} }} < \infty $ for all $t\ge t_{0}$.}
\section{Problem Statement}

A perturbed linear dynamical system is considered:
\begin{equation}\label{eq1}
\begin{array}{l}
y\left( t \right) = \frac{{Z\left( {\theta {\rm{,\;}}s} \right)}}{{R\left( {\theta {\rm{,\; }}s} \right)}}\left[ {u\left( t \right){\rm{ + }}f\left( t \right)} \right]{\rm{,}}\\
Z\left( {\theta {\rm{,\;}}s} \right) = {b_{m - 1}}{s^{m - 1}} + {b_{m - 2}}{s^{m - 2}} +  \ldots  + {b_0}{\rm{,}}\\
R\left( {\theta {\rm{,\;}}s} \right) = {s^n} + {a_{n - 1}}{s^{n - 1}} +  \ldots  + {a_0}{\rm{,}}
\end{array}
\end{equation}
{where $y\left( t \right) \in \mathbb{R}$ is a measurable output, $u\left( t \right) \in \mathbb{R}$ stands for a control signal, $f\left( t \right) \in \mathbb{R}$ denotes an unknown disturbance, \linebreak $\theta  \in {\mathbb{R}^{2n}}$ is a vector of all unknown parameters, $s{\rm{:}} = {\textstyle{d \over {dt}}}$.} {Without loss of generality we assume that the degree of $Z\left(\theta,\;s\right)$ is {\emph{unknown}} and consider the worst case $m = n-1$ throught the paper. If $Z\left( {\theta {\rm{,\;}}s} \right)$ is of known degree $m < n - 1$, then the coefficients ${b_i}{\rm{,\;}}i =\linebreak = n - 1, \ldots {{,}}{ m  +1}$ are equalled to zero and many straightforward simplifications are possible in all further derivations.}

Considering the control, output and disturbance signals, the following assumption is adopted.

\begin{assumption}
The signals $u\left( t \right)$, $y\left( t \right)$ and $f\left( t \right)$ are bounded.
\end{assumption}

To formulate the goal of the study, the parametrization \eqref{eq1} in the form of a LRE is introduced in {full accordance with \cite[p.~49]{b1}:}
\begin{equation}\label{eq2}
z\left( t \right) = {\varphi ^{\top}}\left( t \right)\theta  + w\left( t \right){\rm{,}}
\end{equation}
where
\begin{equation*}
\begin{array}{c}
z\left( t \right) = {\textstyle{{{s^n}} \over {\Lambda \left( s \right)}}}y\left( t \right){\rm{,}}\\
\varphi \left( t \right) = {{\begin{bmatrix}
{ - {\textstyle{{\lambda _{n - 1}^{\top}\left( s \right)} \over {\Lambda \left( s \right)}}}y\left( t \right)}&{{\textstyle{{\lambda _{n - 1}^{\top}\left( s \right)} \over {\Lambda \left( s \right)}}}u\left( t \right)}
\end{bmatrix}}^{\top}}{\rm{, }}\\
w\left( t \right) = {\begin{bmatrix}
{{b_{n - 1}}}&{{b_{n - 2}}}& \ldots &{{b_0}}
\end{bmatrix}}{\textstyle{{\lambda _{n - 1}\left( s \right)} \over {\Lambda \left( s \right)}}}f\left( t \right){\rm{,}}\\
\lambda _{n - 1}^{\top}\left( s \right) = {\begin{bmatrix}
{{s^{n - 1}}}& \cdots &s&1
\end{bmatrix}}{\rm{,\;}}\\\theta  = {{\begin{bmatrix}
{{a_{n - 1}}}&{{a_{n - 2}}}& \cdots &{{b_0}}
\end{bmatrix}}^{\top}}
\end{array}
\end{equation*}
{and $\varphi \left( t \right) \in {\mathbb{R}^{2n}}$ denotes a measurable regressor, $\Lambda \left( s \right)$ is a monic Hurwitz polynomial of the order $n$}.

\textbf{Goal:} to design an estimation law, which ensures that the following equality holds:
\begin{equation}\label{eq4}
\mathop {{\rm{lim}}}\limits_{t \to \infty } \left| {{{\tilde \theta }_i}\left( t \right)} \right| = \mathop {{\rm{lim}}}\limits_{t \to \infty } \left| {{{\hat \theta }_i}\left( t \right) - {\theta _i}} \right| = 0{\rm{\;}}\forall i = 1, \ldots ,2n{\rm{,}}
\end{equation}
where ${\hat \theta _i}\left( t \right)$ is an estimate of the $i^{th}$ unknown parameter.

\begin{remark} %remark1
The above-stated problem \eqref{eq4} is well-studied and solved as far as literature on the offline Identification of Continuous-time Models from Sampled Data \cite{b17, b18, b19} and finite-frequency identification theory \cite{b21} is concerned. {At the same time, to the best of authors’ knowledge, in the literature on the online adaptive control \cite{b1, b2, b3, b4, b5, b6, b7, b8, b9, b10, b11, b12, b13} only P-LS estimation law can provide achievement of the goal \eqref{eq4} under properly PE and independece conditions.} Unlike \cite{b17} and \cite{b21}, in this study the problem \eqref{eq4} is solved online in truly continuous-time and in time domain. {In comparision with P-LS, the proposed solution ensures the novel important properties \textbf{P1}-\textbf{P3}.}
In contrast to the methods from \cite{b3, b11}, in this paper the problem \eqref{eq4} is solved without: ({i}) parameterisation of the disturbance $f\left( t \right)$ into the form of a linear regression and ({ii}) the need to identify the extended vector of unknown parameters.

\end{remark}

\section{The overview of P-LS and IV approaches}

The aim of this section is to {show disadvanteges of P-LS estimator, which provides the stated goal achievement under some restrictions.} {For this purpose the standard P-LS estimation law is introduced:
\begin{equation}\label{eq5}
\begin{array}{l}
\dot {\hat \theta} \left( t \right) =  - \Gamma \left( t \right)\varphi \left( t \right)\left( {{\varphi ^{\top}}\left( t \right)\hat \theta \left( t \right)}-z\left( t \right)  \right){\rm{,\;}}\hat \theta \left( {{t_0}} \right) = {{\hat \theta }_0}{\rm{,}}\\
\dot \Gamma \left( t \right) =  - \Gamma \left( t \right)\varphi \left( t \right){\varphi ^{\top}}\left( t \right)\Gamma \left( t \right){\rm{,\;}}\Gamma \left( {{t_0}} \right) = {\Gamma _0}.
\end{array}
\end{equation}}

{According to simple transformations \cite[p.~50]{b23} and slight abuse of notation of left inversion, the error $\tilde{\theta}(t)$ for \eqref{eq5} obeys the relation:
\begin{equation}\label{eqR}
\tilde \theta_{i} \left( t \right) = \frac{{{\Gamma _0^{ - 1}\tilde \theta \left( {{t_0}} \right) + \int\limits_{{t_0}}^t {\varphi \left( s \right)w\left( s \right)ds} }}}{{{\Gamma _0^{ - 1} + \int\limits_{{t_0}}^t {\varphi \left( s \right){\varphi ^{\top}}\left( s \right)ds}}}}.
\end{equation}}

If $\varphi  \in {\rm{PE}}$, then the following limit holds:
\begin{equation*}
\mathop {{\rm{lim}}}\limits_{t \to \infty } \int\limits_{{t_0}}^t {\varphi \left( s \right){\varphi ^{\top}}\left( s \right)ds}  = \infty {\rm{,}}
\end{equation*}
and, consequently the estimation law \eqref{eq5} ensures that the goal \eqref{eq4} is achieved if $\Gamma_0=\Gamma^{\top}_0 > 0$ and the elements of the regressor $\varphi \left( t \right)$ and perturbation $w\left( t \right)$ are independent{\cite{Fr}}, \emph{i.e.}:
\begin{equation}\label{eq6}
 \forall t \ge {t_0}{\rm{\;}}\left| {\int\limits_{{t_0}}^t {{\varphi _i}\left( s \right)w\left( s \right)ds} } \right| \le c < \infty {\rm{\;}}\forall i = 1, \ldots ,2n{\rm{,}}
\end{equation}
where $c > 0$ is an arbitrary constant.

{Now we are in position to demonstrate that \eqref{eq6} is not met even under a restrictive assumption that $u(t)$ and $f(t)$ can be represented as multiharmonic signals.}

\begin{proposition} Let assumption 1 be met, $u\left( t \right), f\left( t \right)$ be defined as:
\begin{equation}\label{eq3}
\begin{array}{l}
u\left( t \right) = \sum\limits_{k = 1}^n {{\rho _k}{\rm{sin}}\left( {\omega _k^ut} \right)} {\rm{,\;}}\omega _i^u \ne \omega _j^u{\rm{\;}}\forall i \ne j{\rm{,}}\\
f\left( t \right) = \sum\limits_{k = 1}^{{n_f}} {{\delta _k}{\rm{sin}}\left( {\omega _k^ft} \right)} {\rm{,\;}}\omega _i^u \ne \omega _j^f{\rm{\;}}\forall i, j.
\end{array}
\end{equation}

Then there exists $i \in \left\{ {1, \ldots ,2n} \right\} $ such that:
\begin{equation*}
\mathop {{\rm{lim}}}\limits_{t \to \infty } {\rm{ }}\left| {\int\limits_{{t_0}}^t {{\varphi _i}\left( s \right)w\left( s \right)ds} } \right| = \infty .
\end{equation*}    

Proof of proposition 1 is given in Supplementary material \cite{Supp}.
\end{proposition}

Thus, since the signals ${\varphi _i}\left( t \right)$ as well as the perturbation $w\left( t \right)$ are generated with the use of the disturbance $f\left( t \right)$, then under proper assumption \eqref{eq3}, even if there are no common frequencies in $u\left(t\right)$ and $f\left(t\right)$, an element of the regressor exists that violates the condition \eqref{eq6}. Therefore, in case $\varphi  \in {\rm{PE}}$, the law \eqref{eq5} without {additional restrictions on the growing rates of integrals in the numerator and denomirator of \eqref{eqR}} guarantees only boundedness of ${\tilde \theta _i}\left( t \right)$.

To eliminate the dependence between the perturbation and the regressor elements, and hence to ensure that the goal \eqref{eq4} is achieved, in Identification of Continuous-time Models from Sampled Data theory \cite{b17, b18, b19} the method of \emph{instrumental variables} \cite{b15, b16} is widely applied. According to this approach, an instrumental variable $\zeta \left( t \right) \in {\mathbb{R}^{2n}}$ is introduced into consideration such that:
\begin{equation}\label{eq12}
{{\int\limits_{{t_0}}^t {\zeta \left( s \right){\varphi ^{\top}}\left( s \right)ds} } } \to \infty{\rm{\;as\;}}t \to \infty 
\end{equation}
and
\begin{equation}\label{eq13}
 \forall t \ge {t_0}{\rm{\;}}\left| {\int\limits_{{t_0}}^t {{\zeta _i}\left( s \right)w\left( s \right)ds} } \right| \le c < \infty {\rm{\;}}\forall i = 1, \ldots ,2n.
\end{equation}

{This allows one to introduce the IV based P-LS estimator:
\begin{equation}\label{eq14}
\begin{array}{l}
\dot {\hat \theta} \left( t \right) =  - \Gamma \left( t \right)\zeta \left( t \right)\left( {{\varphi ^{\top}}\left( t \right)\hat \theta \left( t \right)}-z\left( t \right) \right){\rm{,\;}}\hat \theta \left( {{t_0}} \right) = {{\hat \theta }_0}{\rm{,}}\\
\dot \Gamma \left( t \right) =  - \Gamma \left( t \right)\zeta \left( t \right){\varphi ^{\top}}\left( t \right)\Gamma \left( t \right){\rm{,\;}}\Gamma \left( {{t_0}} \right) = {\Gamma _0}{\rm ,}
\end{array}
\end{equation}
which provides for $\tilde{\theta}(t)$ the relation:
\begin{equation}\label{eqS}
\tilde \theta \left( t \right) = \frac{{\Gamma _0^{ - 1}\tilde \theta \left( {{t_0}} \right) + \int\limits_{{t_0}}^t {\zeta \left( s \right)w\left( s \right)ds} }}{{\Gamma _0^{ - 1} + \int\limits_{{t_0}}^t {\zeta \left( s \right){\varphi ^{\top}}\left( s \right)ds} }}.
\end{equation}}

{The following conclusions can be made from \eqref{eqS}:
\begin{enumerate}
    \item [\textbf{C1)}] to provide asymptotic convergence, IV based P-LS \eqref{eqS} needs to satisfy not only \eqref{eq12} and \eqref{eq13}, but also a hardly validated invertability condition for some $\gamma>0$ :
    \begin{equation*}
\left|{\rm{det}}\left\{ {\Gamma _0^{ - 1} + \int\limits_{{t_0}}^t {\zeta \left( s \right){\varphi ^{\top}}\left( s \right)ds} } \right\}\right| \ge \gamma  > 0{\rm{\;}}\forall t  \ge {t_0}{\rm ,}
\end{equation*}
    \item[\textbf{C2)}] the condition \eqref{eq12} can be considered as equivalent to the PE one, which is rather stringent for practical applications; 
        \item[\textbf{C3)}] the convergence rate of $\tilde{\theta}\left(t\right)$ does not exceed the asymptotic one even if $w\left(t\right)=0$ and condition \eqref{eq12} is satisfied. 
\end{enumerate}}

{Therefore, an actual problem is to design an estimation law that guarantees the achievement of the goal \eqref{eq4} under weaker conditions and ensures improved convergence rate in some cases.} In the next section of this study such a law is developed on the basis of the dynamic regressor extension and mixing procedure \cite{b4}, the above-considered method of instrumental variables \cite{b15, b16, b17, b18, b19} and the filtering with averaging \cite{b20}.

\section{Main Result}

The description of the main result of this study is divided into {five parts}. Subsection A introduce the {novel continuous-time online scheme} to generate instrumental variables. In Subsection B a modified dynamic regressor extension and mixing scheme is proposed on the basis of the idea of regressor extension with the help of instrumental variable and filtering with averaging. According to this procedure, the regression equation \eqref{eq2} is reduced to a set of scalar ones ${{\cal Y}_i}\left( t \right) = \Delta \left( t \right){\theta _i} +\linebreak+ {{\cal W}_i}\left( t \right)$, in which the perturbation ${{\cal W}_i}\left( t \right)$ is asymptotically decreasing if the condition \eqref{eq13} is satisfied. The gradient estimation law is then applied to identify the parameters of the obtained regression equations. {In subsection C a conservative convergence condition for the proposed identifier is described. Further subsection D shows that the proposed estimation law actually ensures assymptotic convergence when weaker conditions are met. In subsection E obtained results are extended to the problem of identification in the closed loop.} Some basic definitions from harmonic analysis (see {Supplementary materials \cite{Supp}}) are used to derive the results of this section.

\subsection{Dynamic generation of instrumental variable}

In accordance with Section III, the instrumental variable needs to meet the conditions \eqref{eq12} and \eqref{eq13} to ensure achievement of the goal \eqref{eq4} when P-LS is applied. In the literature on Identification of Continuous-time Models from Sampled Data \cite{b16, b17}, about four methods of instrumental variable generation have been proposed to meet similar conditions. {However, none of them has been applied to the considered \emph{online} identification problem in continuous time.} In this study, it is proposed to form $\zeta \left( t \right)$ with the help of the following auxiliary filter:
\begin{equation}\label{eq18}
\begin{array}{l}
{y_{iv}}\left( t \right) = \frac{{Z\left( {{\theta _{iv}}{\rm{,\;}}s} \right)}}{{R\left( {{\theta _{iv}}{\rm{,\;}}s} \right)}}u\left( t \right){\rm{,}}\\
\zeta \left( t \right) = {{\begin{bmatrix}
{ - {\textstyle{{\lambda _{n - 1}^{\top}\left( s \right)} \over {\Lambda \left( s \right)}}}{y_{iv}}\left( t \right)}&{{\textstyle{{\lambda _{n - 1}^{\top}\left( s \right)} \over {\Lambda \left( s \right)}}}u\left( t \right)}
\end{bmatrix}}^{\top}}{\rm{, }}
\end{array}
\end{equation}
where ${\theta _{iv}}$ are known parameters of the instrumental model such that $R\left( {{\theta _{iv}}{\rm{,\;}}s} \right)$ is a Hurwitz polynomial.

The following propositions hold for the instrumental variable \eqref{eq18}.
\begin{proposition} {{Define
\begin{equation*}
    \begin{array}{l}
        H\left( s \right) = {{\begin{bmatrix}
{ - {\textstyle{{\lambda _{n - 1}^{\top}\left( s \right)Z\left( {\theta {\rm{,\;}}s} \right)} \over {\Lambda \left( s \right)R\left( {\theta {\rm{,\;}}s} \right)}}}}&{{\textstyle{{\lambda _{n - 1}^{\top}\left( s \right)} \over {\Lambda \left( s \right)}}}}
\end{bmatrix}}^{\top}}{\rm{,\;}}\\M\left( s \right) = {\begin{bmatrix}
{G\left( s \right){I_n}}&0\\
0&{{I_n}}
\end{bmatrix}}{\rm{,}}
\end{array}
\end{equation*}
where $G\left( s \right)$ is such that ${\textstyle{{G\left( s \right)Z\left( {\theta {\rm{,\;}}s} \right)} \over {R\left( {\theta {\rm{,\;}}s} \right)}}} = {\textstyle{{Z\left( {{\theta _{iv}}{\rm{,\;}}s} \right)} \over {R\left( {{\theta _{iv}}{\rm{,\;}}s} \right)}}}$.}}

{Assume that:}
    \begin{enumerate}
    \item [\textbf{1)}] {$u\left( t \right)$ is stationary,}
        \item [\textbf{2)}] $H\left( {j{\omega _1}} \right){\rm{,\;}} \ldots {\rm{,\;}}H\left( {j{\omega _{2n}}} \right)$ are linearly independent in ${\mathbb{C}^{2n}}$ for all ${\omega _1}{\rm{,\;}} \ldots {\rm{,\;}}{\omega _{2n}}$ such that ${\omega _i} \ne {\omega _k}$ for $i \ne k$;
        \item [\textbf{3)}] $rank\left\{ {\left[ {M\left( {j{\omega _1}} \right){\rm{,\;}} \ldots {\rm{,\;}}M\left( {j{\omega _{2n}}} \right)} \right]} \right\} = 2n$ for all ${\omega _1}{\rm{,}} \ldots{\rm{,\;}} {\omega _{2n}}$ such that ${\omega _i} \ne {\omega _k}$ for $i \ne k.$
    \end{enumerate}

Then there exist $T > 0$ and $\alpha  > 0$ such that
\begin{equation}\label{eq19}
\left| {{\rm{det}}\left\{ {\int\limits_t^{t + T} {\zeta \left( \tau  \right){\varphi ^{\top}}\left( \tau  \right)d\tau } } \right\}} \right| \ge \alpha  > 0,\;\forall t \ge {t_0}{\rm{,}}
\end{equation}
if and only if $u\left( t \right)$ is sufficiently rich of order $2n$.

Proof of proposition 2 is presented in Supplementary material \cite{Supp}.
\end{proposition}

\begin{proposition} 
Let assumption 1 be met and $u\left( t \right),\;f\left( t \right)$ be defined according to \eqref{eq3}. Then the instrumental variable \eqref{eq18} meets the condition \eqref{eq13}.

Taking into consideration the proof of proposition 1, it immediately follows that proposition 3 holds.
\end{proposition} 

Proposition 2 is an analog of the classical result \cite{b22}, in which necessary and sufficient conditions for the regressor $\varphi \left( t \right)$ to be persistently exciting were formulated. In comparison with the condition $\varphi  \in {\rm{PE}}$, the existence of a controller $G\left( s \right)$ providing ${\textstyle{{G\left( s \right)Z\left( {\theta {\rm{,\;}}s} \right)} \over {R\left( {\theta {\rm{,\;}}s} \right)}}} = {\textstyle{{Z\left( {{\theta _{iv}}{\rm{,\;}}s} \right)} \over {R\left( {{\theta _{iv}}{\rm{,\;}}s} \right)}}},$ as well as the independence of $M\left( {j{\omega _1}} \right){\rm{,\;}} \ldots {\rm{,\;}}M\left( {j{\omega _{2n}}} \right)$ are additionally required to satisfy the inequality \eqref{eq19} and then \eqref{eq12}. These new conditions are not restrictive. \textcolor{black}{On the other hand, in proposition 3 the \emph{sufficient} conditions to satisfy \eqref{eq13} are formulated.}

{It should be specially noted that, to the best of authors' knowledge, in this study the properties \eqref{eq13} and \eqref{eq19} were proved for the instrumental variable \eqref{eq18} in the considered continuous-time online parameter estimation task for the first time. Previously analogous conditions were well known only for the offline or discrete time identifications cases \cite{b16, b17, b18, b19}.} The fact that the premises of propositions 1 and 2 hold allows one to \emph{bona fide} use the instrumental variable \eqref{eq18} in  continuous time P-LS law \eqref{eq14} and for the further design of the novel estimation law.

\subsection{IV based DREM}

In recent years, the procedure of dynamic regressor extension and mixing (DREM) has attracted great attention \cite{b4}, as it allows one to design the estimation law with relaxed convergence conditions in comparison with the classical approaches and provides an improved transient quality of the parameter estimates. DREM consists of two main stages: dynamic regressor extension and mixing. In the first step, the original equation is transformed by linear operations and dynamical operators into a regression equation with a new regressor, which is a square matrix. At the second stage, an algebraic transformation is applied to transform the equation obtained at the extension stage into a set of scalar independent from each other equations with respect to the components of the vector $ \theta$. In this study, it is proposed to use a novel mixing scheme in DREM procedure obtained as a combination of the sliding window extension scheme \cite[Lemma 1]{b14} with the instrumental variables approach \cite{b15, b16, b17, b18, b19}:
\renewcommand{\theequation}{\arabic{equation}a}
\begin{equation}\label{eq21a}
\begin{array}{l}
\dot \vartheta \left( t \right) = \zeta \left( t \right)z\left( t \right) - \zeta \left( {t - T} \right)z\left( {t - T} \right){\rm{,\;}}\\
\dot \psi \left( t \right) = \zeta \left( t \right){\varphi ^{\top}}\left( t \right) - \zeta \left( {t - T} \right){\varphi ^{\top}}\left( {t - T} \right){\rm{,\;}}\\
\vartheta \left( {{t_0}} \right) = {0_{2n}}{\rm{,}}\;\psi \left( {{t_0}} \right) = {0_{2n \times 2n}}{\rm{,}}
\end{array}
\end{equation}
and filtering with averaging \cite{b20}:
\setcounter{equation}{13}
\renewcommand{\theequation}{\arabic{equation}b}
\begin{equation}\label{eq21b}
\begin{array}{l}
\dot Y\left( t \right) =  - \frac{1}{{F\left( t \right)}}\dot F\left( t \right)\left( {Y\left( t \right) - \vartheta \left( t \right)} \right){\rm{,\;}}\\
\dot \Phi \left( t \right) =  - \frac{1}{{F\left( t \right)}}\dot F\left( t \right)\left( {\Phi \left( t \right) - \psi \left( t \right)} \right){\rm{,\;}}\\
\dot F\left( t \right) = p{t^{p - 1}}{\rm{,\;}}\\ 
Y\left( {{t_0}} \right) = {0_{2n}}{\rm{,\;}}\Phi \left( {{t_0}} \right) = {0_{2n \times 2n}}{\rm{,\;}}F\left( {{t_0}} \right) = {F_0},
\end{array}
\end{equation}
where $T > 0$ denotes a sliding window width, {$p \ge 1,{\rm{\;}}{F_0} > t_{0}^p$} 
 stand for the filter parameters.
\renewcommand{\theequation}{\arabic{equation}}

\begin{proposition} Let $\theta  = const$, then the signals $Y\left( t \right)$ and $\Phi \left( t \right)$ meet:
    \begin{equation}\label{eq22}
Y\left( t \right) = \Phi \left( t \right)\theta  + W\left( t \right){\rm{,}}
\end{equation}
where the new disturbance $W\left( t \right)$ satisfies the equations:
    \begin{equation*}
\begin{array}{l}
\dot W\left( t \right) =  - \frac{1}{{F\left( t \right)}}\dot F\left( t \right)\left( {W\left( t \right) - \varepsilon \left( t \right)} \right){\rm{,\;}}W\left( {{t_0}} \right) = {0_{2n}}{\rm ,}\\
\dot \varepsilon \left( t \right) = \zeta \left( t \right)w\left( t \right) - \zeta \left( {t - T} \right)w\left( {t - T} \right){\rm{,\;}}\varepsilon \left( {{t_0}} \right) = {0_{2n}}.\\
\end{array}
\end{equation*}

Proof of proposition is given in Supplementary material \cite{Supp}.
\end{proposition}

Considering the mixing step, the left- and right-hand sides of equation \eqref{eq22} are multiplied by an adjoint matrix ${\rm{adj}}\left\{ {\Phi \left( t \right)} \right\}$, which, owing to ${\rm{adj}}\left\{ {\Phi \left( t \right)} \right\}\Phi \left( t \right) =\linebreak= {\rm{det}}\left\{ {\Phi \left( t \right)} \right\}{I_{2n \times 2n}}$, allows one to obtain a set of scalar separate regression equations:
    \begin{equation}\label{eq24}
{{\cal Y}_i}\left( t \right) = \Delta \left( t \right){\theta _i} + {{\cal W}_i}\left( t \right){\rm{,}}
\end{equation}
where
\begin{equation*}
    \begin{array}{c}
{\cal Y}\left( t \right){\rm{:}} = {\rm{adj}}\left\{ {\Phi \left( t \right)} \right\}Y\left( t \right){\rm{,\;}}\Delta \left( t \right){\rm{:}} = {\rm{det}}\left\{ {\Phi \left( t \right)} \right\}{\rm{,}}\\
{\cal W}\left( t \right){\rm{:}} = {\rm{adj}}\left\{ {\Phi \left( t \right)} \right\}W\left( t \right){\rm{,}}\\
{\cal Y}\left( t \right) = {{\begin{bmatrix}
{{{\cal Y}_1}\left( t \right)}& \ldots &{{{\cal Y}_{i - 1}}\left( t \right)}&{\begin{array}{*{20}{c}}
 \ldots &{{{\cal Y}_{2n}}\left( t \right)}
\end{array}}
\end{bmatrix}}^{\top}}{\rm{,}}\\
{\cal W}\left( t \right) = {{\begin{bmatrix}
{{{\cal W}_1}\left( t \right)}& \ldots &{{{\cal W}_{i - 1}}\left( t \right)}&{\begin{array}{*{20}{c}}
 \ldots &{{{\cal W}_{2n}}\left( t \right)}
\end{array}}
\end{bmatrix}}^{\top}}.
\end{array}
\end{equation*}

The scheme \eqref{eq21a} in case $\zeta \left( t \right)-\varphi\left(t\right) = 0$ can be reduced to the sliding window extension scheme from \cite{b14}. However, the application of the averaging filters \eqref{eq21b} and choice of the instrumental variable as \eqref{eq18} provide the regressor $\Delta \left( t \right)$ and disturbances ${{\cal W}_i}\left( t \right)$ with the new properties. Using the obtained set of equations \eqref{eq24} the following simple gradient descent estimation law is introduced:
         \begin{equation}\label{eq26}
{\dot {\hat \theta} _i}\left( t \right) =  - \gamma \Delta \left( t \right)\left( {\Delta \left( t \right){{\hat \theta }_i}\left( t \right) - {{\cal Y}_i}\left( t \right)} \right){\rm{,\;}}{\hat \theta _i}\left( {{t_0}} \right) = {\hat \theta _{0i}},
\end{equation}   
where $\gamma > 0$.

{Naturally, in comparison with P-LS \eqref{eq14}, the proposed estimator does not need properly selected initial conditions to ensure bounded parameters estimations, \emph{e.g.} it has been proved recently \cite{b12, b13} that $\tilde \theta \left(t\right)$ is bounded if ${\mathcal{W}}_{i} \in L_2$ for all $i={1,...,2n}$.} {In the next two subsections we present the conservative and relaxed convergence conditions for the law \eqref{eq26}.}

\subsection{Conservative convergence conditions}

{To derive conservative convergence conditions, first of all, the inequalities \eqref{eq13} and \eqref{eq19} are to be reformulated into properties of the obtained scalar regressor $\Delta\left(t\right)$ and perturbation ${\mathcal{W}}_{i}\left(t\right)$.}

\begin{proposition} The following statements hold:
\begin{enumerate}
    \item [\textbf{S1)}] {if for the selected sliding window width $T > 0$ there exists $\alpha > 0$ that makes \eqref{eq19}} true, {then also there exist ${\Delta _{{\rm{LB}}}} > 0$} and {$T_{\Delta}\ge t_{0}+T$} such that
    \begin{equation*}
  \left| {\Delta \left( t \right)} \right| \ge {\Delta _{{\rm{LB}}}} > 0{\rm{\;}}\forall t \ge {T_{\Delta}}{\rm{,}}        
    \end{equation*}
\item [\textbf{S2)}] if assumption 1 is met, then {there exists ${\Delta _{{\rm{UB}}}} > 0$ such that}
    \begin{equation*}
\left| {\Delta \left( t \right)} \right| \le {\Delta _{{\rm{UB}}}}{\rm{\;}}\forall t \ge {t_0}{\rm{,}}       
    \end{equation*}
    \item [\textbf{S3)}] if inequalities \eqref{eq13} hold and assumption 1 is met, then {${{\cal W}_i} \in {L_{l}}$ for any $l \in \left( {1,{\rm{ }}\infty } \right)$} and there exists ${c_{\cal W}} > 0$, such that:
        \begin{equation}\label{eq25}
\left| {{{\cal W}_i}\left( t \right)} \right| \le \frac{{\dot F\left( t \right){c_{\cal W}}}}{{F\left( t \right)}} < \infty .
\end{equation} 
\end{enumerate}

Proof of proposition 5 is presented in Supplementary material \cite{Supp}.
\end{proposition}

Thus, the DREM procedure based on IV approach and filtering with averaging allows one to obtain a set of scalar regression equations \eqref{eq24}, in which the perturbation ${{\cal W}_i}\left( t \right)$ satisfies the condition \eqref{eq25}, i.e., {it asymptotically decreases and is integrable with finite degree greater than one.} The properties obtained in proposition 5 make it possible to prove the convergence of the obtained estimates in the following theorem.

\begin{theorem} The stated goal \eqref{eq4} is achieved if premises \textbf{S1}-\textbf{S3} hold. {Moreover, if aditionally 
\begin{equation}\label{Cond}
\forall t \ge {T} \int\limits_{{\rm{max}}\left\{ {{t_0}{\rm{, }}t - T} \right\}}^t {\zeta_{i} \left( s \right)w\left( s \right)ds}  = 0,{\rm{\;}}\forall i = 1, \ldots ,2n,
\end{equation}
then for all $t \ge T_{\Delta}$ the following upper bound exists:
\begin{equation*}
    \left| {{{\tilde \theta }_i}\left( t \right)} \right| \le {e^{ - \gamma {\Delta _{{\rm{LB}}}}\left( {t - T_{\Delta}} \right)}}\left| {{{\tilde \theta }_i}\left( {{T_{\Delta}}} \right)} \right|{\rm{\;}}\forall i = 1, \ldots ,2n.
\end{equation*}}
Proof of theorem can be found in Supplementary material \cite{Supp}.

\end{theorem}

Therefore, unlike the existing online identification laws \cite{b1, b2, b3, b4, b5, b6, b7, b8, b9, b10, b11, b12, b13, b14}, the proposed one \eqref{eq18} + \eqref{eq21a} + \eqref{eq21b} + \eqref{eq26} is capable of identifying the exact values of the unknown parameters of linear systems affected by external disturbances. {In comparison with throughly discussed IV based P-LS law \eqref{eq14}, the proposed estimator does not need to satisfy the invertability condition and ensures exponential convergence of the parametric error in perturbation-free case.}

The first condition of exact asymptotic convergence \eqref{eq19} is referred to the modified notion of the persistence of excitation, and the second condition \eqref{eq13} requires independence (in the sence of definition 2) of the perturbation $w(t)$ and instrumental variables $\zeta_{i}(t)$ from each other. Particularly, how it can be seen from propositions 1, 3,  the independence requirement \eqref{eq13} is not restrictive and satisfied when, {{\it e.g.},} the control input $u(t)$ and plant perturbation $f(t)$ do not contain common frequencies.

\begin{remark}%remark 2
{It is interesting to see that $w\left(t\right)=0$ is only sufficient condition to satisfy equality \eqref{Cond}, and there exists a ''good choice'' of width $T$, which ensures disturbance annihilation. For example, if $\zeta_{i}\left(t\right)=1$, $w\left(t\right)={\rm{sin}}\left({\omega}t\right)$ and $T = {\textstyle{{2\pi } \over \omega }}$, then \eqref{Cond} is met.}
\end{remark}

\begin{remark}%remark 3
{The control and perturbation signals from the special class \eqref{eq3} are introduced in this study only to show that the parametric convergence conditions \eqref{eq13} and \eqref{eq19} can be met. According to the proof of Proposition 5 and Theorem 1, the proposed solution ensures parametric convergence under any other control and perturbation signals, which application results in fulfilment of \eqref{eq13} and \eqref{eq19}. For example, Proposition 5 premises do not exclude the situation when \eqref{eq19} is met for a non-stationary signal $u(t)$.}
\end{remark}

\subsection{Relaxed convergence conditions}

{For the most practical applications the system input $u\left(t\right)$ is not necessarily sufficiently rich, and consequently inequality \eqref{eq19} becomes restrictive without artifically dither signal injection. In the next theorem the parameter convergence is established for the law \eqref{eq26} under weaker conditions.}
{
\begin{theorem} If premise \textbf{S3} holds and $\Delta \notin L_{2}$, then the stated goal \eqref{eq4} is achieved.
\end{theorem}}
\proof{To prove the theorem, first of all, the derivative of a quadratic form $V\left(t\right) = {\textstyle{1 \over 2}}\tilde \theta _i^2\left(t\right)$ is obtained as follows:
\begin{displaymath}
\begin{array}{l}
\dot V\left(t\right) =  - \gamma {\Delta ^2\left(t\right)}\tilde \theta _i^2\left(t\right) + \gamma {\tilde \theta _i}\Delta\left(t\right) {{\cal W}_i}\left(t\right) \le \\ \le  - \gamma {\Delta ^2}\left(t\right)\tilde \theta _i^2\left(t\right) + \gamma \left| {{{\tilde \theta }_i}}\left(t\right) \right|\left| \Delta\left(t\right)  \right|\left| {{{\cal W}_i}}\left(t\right) \right|.
\end{array}
\end{displaymath}}
{Having applied the inequality $ab \le 2ab \le \delta {a^2} + {\delta ^{ - 1}}{b^2}$, the following upper bound is obtained:
\begin{displaymath}\begin{array}{l}
\dot V\left(t\right) \le  - \gamma {\Delta ^2\left(t\right)}\left( {1 - \delta } \right)\tilde \theta _i^2\left(t\right) + \gamma {\delta ^{ - 1}}{\cal W}_i^2\left(t\right) = \\=  - \gamma {\Delta ^2}\left(t\right)\left( {1 - \delta } \right)V\left(t\right) + \gamma {\delta ^{ - 1}}{\cal W}_i^2\left(t\right){\rm{,}}
\end{array}
\end{displaymath}
where $\delta  \in \left( {0,{\rm{\;1}}} \right)$. }

{Now we recall the usefull results of \cite[Lemma 3.1]{Ara}.}
\begin{lemma} {Consider the scalar system definded by
\begin{displaymath}
\dot x\left( t \right) =  - {a^2}\left( t \right)x\left( t \right) + b\left( t \right){\rm{,\;}}x\left( {{t_0}} \right) = {x_0}{\rm ,}
\end{displaymath}
where $x\left(t\right)\in\mathbb{R}$, $a,\;b{\rm{\;:\;}}\mathbb{R}_{+}\mapsto\mathbb{R}$ are piecewise continuous bounded functions. If $a\notin L_{2}$ and $b\in L_{1}$ then $\mathop {{\rm{lim}}}\limits_{t \to \infty } x\left( t \right) = 0$.}

{Proof of Lemma is presented in \cite[Section 3.A.1]{Ara}.}
\end{lemma}

{From \textbf{S3} it is obvious that $\mathcal{W}^2_{i}\in L_{1}$. Therefore,  if aditionally \linebreak $\Delta\notin L_2$, then all premises of Lemma 1 are met and, owing to Comparison Lemma \cite[Lemma 3.4]{Khalil}, the stated goal is achieved.}\endproof

\begin{remark}%remark 4
{Theorem 2 improves the results of \cite[Proposition 1]{b12}, \cite[Theorem 3]{b13}, in which only boundedness of $\tilde \theta_{i}\left(t\right)$ is shown for the case $\mathcal{W}_{i}\in L_{2}$ and $\Delta \notin L_{2}$. Moreover, in contrast to standard DREM, the novel IV based DREM \eqref{eq21a} + \eqref{eq21b} + \eqref{eq24} strictly provides fulfillment of the requirement $\mathcal{W}_{i}\in L_{2}$.} 
\end{remark}

{For the following signals all requirements from theorem 2 are met:
\begin{displaymath}
{\rm{\Delta }}\left( {{t}}  \right){\rm{:}} = {\textstyle{1 \over {\sqrt {t + 1} }}}{\rm{,\;}}{\mathcal{W}_i}\left( t \right){\rm{:}} = {\textstyle{1 \over {t + 1}}}{\rm{}}
\end{displaymath}
and consequently the interpretation of convergence conditions from theorem 2 is that the perturbation ${\mathcal{W}}_{i}\left(t\right)$ needs to converge to zero more rapidly in comparison with $\Delta\left(t\right)$.} 

{In theorem 1 the convergence is proved under the requirement that the regressor is bounded away from zero (see \textbf{S1} from proposition 5), which, according to proposition 2, relates to sufficiently rich condition that is restrictive for applications. In theorem 2 the convergence is shown without afromentioned requirement as, for instance, there exists a signal ${\rm{\Delta }}\left( {{t}}  \right){\rm{:}} = {\textstyle{1 \over {\sqrt {t + 1} }}}$, which is not globally bounded away from zero, but also it is not in $L_{2}$. More preciciely, the condition \textbf{S1} means that the minimum eigenvalue of $\Phi\left(t\right)$ is bounded away from zero, while the condition $\Delta \notin L_{2}$ means that all eigenvalues of $\Phi\left(t\right)$ are not square integrable. The result of theorem 2 is important for practical scenarios as it shows that estimates $\hat\theta\left(t\right)$ could converge to $\theta$ without restrictive stationarity and richness assumptions. Unfortunally, the significance of such relaxation is slightly mitigated as now we do not provide \emph{bona fide} answer to the following question:}

{\textbf{Q. }What signals $u\left(t\right)$ and $f\left(t\right)$ do ensure the fulfillment of both the independence requirement \eqref{eq13} and condition $\Delta\notin L_{2}$? }

\subsection{Extension to identification in closed loop}
The above-proposed estimation law is applicable to the identification of system parameters in an open loop. In case of the closed-loop system, in which the controller is feedback-based one, the control signal $u(t)$ depends on the disturbance, which results in the common frequencies in their spectra and the violation of condition \eqref{eq13} in case the instrumental model \eqref{eq18} is used. In this subsection a method to choose an instrumental model is proposed, which \textcolor{black}{provides condition} \eqref{eq13} satisfaction in case of identification in closed loop.

It is assumed that the control signal $u\left( t \right)$ is formed as follows:
\begin{equation}\label{1N}
u\left( t \right) = \frac{{{P_y}\left( {\kappa {\rm{,\;}}s} \right)}}{{{Q_y}\left( {\kappa {\rm{,\;}}s} \right)}}y\left( t \right) + \frac{{{P_r}\left( {\kappa {\rm{,\;}}s} \right)}}{{{Q_r}\left( {\kappa {\rm{,\;}}s} \right)}}r\left( t \right){\rm{,}}    
\end{equation}
where $\kappa  \in {\mathbb{R}^{{n_\kappa }}}$ are known time-invariant parameters of the control law, $r\left( t \right){\in \mathbb{R}}$ stands for the reference signal, ${m_y} \le {n_y}$ and ${m_r} \le {n_r}$ are orders of the couples of polynomials ${P_y}\left( {\kappa {\rm{,\;}}s} \right){\rm{,\;}}{Q_y}\left( {\kappa {\rm{,\;}}s} \right)$ and ${P_r}\left( {\kappa {\rm{,\;}}s} \right){\rm{,\;}}{Q_r}\left( {\kappa {\rm{,\;}}s} \right)$, respectively.

The instrumental variable $\zeta \left( t \right)$ is proposed to be chosen as:
\begin{equation}\label{3N}
\begin{array}{l}
{y_{iv}}\left( t \right) = \frac{{Z\left( {{\theta _{iv}}{\rm{,\;}}s} \right)}}{{R\left( {{\theta _{iv}}{\rm{,\;}}s} \right)}}{u_{iv}}\left( t \right){\rm{,\;}}\\
{u_{iv}}\left( t \right) = \frac{{{P_y}\left( {\kappa {\rm{,\;}}s} \right)}}{{{Q_y}\left( {\kappa {\rm{,\;}}s} \right)}}{y_{iv}}\left( t \right) + \frac{{{P_r}\left( {\kappa {\rm{,\;}}s} \right)}}{{{Q_r}\left( {\kappa {\rm{,\;}}s} \right)}}r\left( t \right){\rm{,}}\\
\zeta \left( t \right) = {{\begin{bmatrix}
{ - {\textstyle{{\lambda _{n - 1}^{\top}\left( s \right)} \over {\Lambda \left( s \right)}}}{y_{iv}}\left( t \right)}&{{\textstyle{{\lambda _{n - 1}^{\top}\left( s \right)} \over {\Lambda \left( s \right)}}}{u_{iv}}\left( t \right)}
\end{bmatrix}}^{\top}}{\rm{,\;}}
\end{array}    
\end{equation}
where ${\theta _{iv}}$ are known parameters of the instrumental model such that the closed-loop model \eqref{3N} is stable.

In contrast to \eqref{eq18}, the instrumental variable \eqref{3N} does not depend on the control signal $u(t)$, which allows one to ensure that the condition \eqref{eq13} is met when the system \eqref{eq1} is inside a closed loop. Further the analogs of propositions 2 and 3 are formulated for the closed-loop system \eqref{eq1} + \eqref{1N}.

\begin{proposition}%proposition 6
{Define
\begin{displaymath}
\begin{array}{l}
H\left( s \right) = {{\begin{bmatrix}
{ - {\textstyle{{\lambda _{n - 1}^{\top}\left( s \right){W_{cl}}\left( {\theta {\rm{,\;}}s} \right)Z\left( {\theta {\rm{,\;}}s} \right)} \over {\Lambda \left( s \right)}}}}&{{\textstyle{{\lambda _{n - 1}^{\top}\left( s \right){W_{cl}}\left( {\theta {\rm{,\;}}s} \right)R\left( {\theta {\rm{,\;}}s} \right)} \over {\Lambda \left( s \right)}}}}
\end{bmatrix}}^{\top}}{\rm{,\;}}\\
M\left( s \right) = {\begin{bmatrix}
{{G_1}\left( s \right)}&0\\
0&{{G_2}\left( s \right)}
\end{bmatrix}} {\rm{,}}\\
{W_{cl}}\left( {\theta {\rm{, }}s} \right) = {\textstyle{{{P_r}\left( {\kappa {\rm{, }}s} \right){Q_y}\left( {\kappa {\rm{,\;}}s} \right)} \over {{Q_r}\left( {\kappa {\rm{,\;}}s} \right)\left[ {{Q_y}\left( {\kappa {\rm{,\;}}s} \right)R\left( {\theta {\rm{,\;}}s} \right) - Z\left( {\theta {\rm{,\;}}s} \right){P_y}\left( {\kappa {\rm{,\;}}s} \right)} \right]}}}{\rm{,\;}}
\end{array}    
\end{displaymath}
where ${G_1}\left( s \right){\rm{,\;}}{G_2}\left( s \right)$ are such that
\begin{displaymath}
\begin{array}{l}
{G_1}\left( s \right)\lambda _{n - 1}\left( s \right){W_{cl}}\left( {\theta {\rm{,\;}}s} \right)Z\left( {\theta {\rm{,\;}}s} \right) = \quad\quad \quad\quad\quad\quad\quad\quad \quad\quad\quad\quad\hfill\\\quad\quad\quad\quad\quad\quad\quad\quad\hfill
=\lambda _{n - 1}\left( s \right){W_{cl}}\left( {{\theta _{iv}}{\rm{,\;}}s} \right)Z\left( {{\theta _{iv}}{\rm{,\;}}s} \right){\rm{,}}\hfill\\
{G_2}\left( s \right)\lambda _{n - 1}\left( s \right){W_{cl}}\left( {\theta {\rm{,\;}}s} \right)R\left( {\theta {\rm{,\;}}s} \right) = \\\quad\quad\quad\quad\quad\quad\quad\quad\hfill
=\lambda _{n - 1}\left( s \right){W_{cl}}\left( {{\theta _{iv}}{\rm{,\;}}s} \right)R\left( {{\theta _{iv}}{\rm{,\;}}s} \right).\hfill
\end{array}    
\end{displaymath}}
{Assume that}
\begin{enumerate}
\item[\textbf{1)}] {$r\left( t \right)$ is stationary,}
    \item[\textbf{2)}] $H\left( {j{\omega _1}} \right){\rm{,\;}} \ldots {\rm{,\;}}H\left( {j{\omega _{2n}}} \right)$ are linearly independent in ${\mathbb{C}^{2n}}$ for all ${\omega _1}{\rm{,\;}} \ldots {\rm{,\;}}{\omega _{2n}}$ such that ${\omega _i} \ne {\omega _k}$ for $i \ne k$;
    \item[\textbf{3)}] $rank\left\{ {\left[ {M\left( {j{\omega _1}} \right){\rm{,\;}} \ldots {\rm{,\;}}M\left( {j{\omega _{2n}}} \right)} \right]} \right\} = 2n$ for all ${\omega _1}{\rm{,\;}} \ldots {\rm{,\;}}{\omega _{2n}}$ such that ${\omega _i} \ne {\omega _k}$ for $i \ne k.$
\end{enumerate}

Then there exist $T > 0$ and $\alpha  > 0$ such that \eqref{eq19} is met if and only if $r\left( t \right)$ is sufficiently rich of order $2n$.

Proof of proposition 6 is presented in Supplementary material \cite{Supp}.   
\end{proposition}

\begin{proposition}%proposition 7
Let Assumption 1 be met and $r\left( t \right)$, $f\left( t \right)$ be defined as:
\begin{equation*}
\begin{array}{l}
r\left( t \right) = \sum\limits_{k = 1}^n {{\rho _k}{\rm{sin}}\left( {\omega _k^rt} \right)} {\rm{,\;}}\omega _i^r \ne \omega _j^r{\rm{\;}}\forall i \ne j{\rm{,}}\\
f\left( t \right) = \sum\limits_{k = 1}^{{n_f}} {{\delta _k}{\rm{sin}}\left( {\omega _k^ft} \right)} {\rm{,\;}}\omega _i^r \ne \omega _j^f{\rm{\;}}\forall i{\rm{,\;}}j{\rm{,}}
\end{array}    
\end{equation*}
then the condition \eqref{eq13} is satisfied.

Proof of this proposition coincides with the one for proposition 3.    
\end{proposition}

Therefore, when the instrumental model is chosen as \eqref{3N} and premises of theorem 1 or theorem 2 are satisfied, the proposed estimation law \eqref{eq21a} + \eqref{eq21b} + \eqref{eq26} ensures exact asymptotic identification of the unknown parameters of system \eqref{eq1} even in a closed-loop case.

\begin{remark}%remark 4
It should be mentioned, that if $Z\left( {\theta {\rm{,\;}}s} \right)$ is stable, then, based on the obtained estimates, the perturbation $f\left( t \right)$ can be approximately estimated through the following simple certainty equivalence adaptive disturbance observer:
\begin{displaymath}
\hat f\left( t \right){\rm{:}} = {\textstyle{{{P_f}\left( s \right)R\left( {\hat \theta {\rm{,\;}}s} \right)} \over {{Q_f}\left( s \right)Z\left( {\hat \theta {\rm{,\;}}s} \right)}}}y\left( t \right) - {\textstyle{{{P_f}\left( s \right)} \over {{Q_f}\left( s \right)}}}u\left( t \right){\rm{,}}    
\end{displaymath}
where ${P_f}\left( s \right){\rm{,\;}}{Q_f}\left( s \right)$ are chosen such that the transfer function ${\textstyle{{{P_f}\left( s \right)R\left( {\theta {\rm{,\;}}s} \right)} \over {{Q_f}\left( s \right)Z\left( {\theta {\rm{,\;}}s} \right)}}}$ is proper and satisfy ${\textstyle{{{P_f}\left( 0 \right)} \over {{Q_f}\left( 0 \right)}}} = 1$.    
\end{remark}

\section{Numerical Simulation}
The below-given linear dynamical system of the second order with the piecewise-constant parameters has been considered:
\begin{equation}\label{eq27}
y\left( t \right) = \left\{ \begin{array}{l}
\frac{{ - 2s - 1}}{{{s^2} + s + 2}}\left[ {u\left( t \right){\rm{ + }}f\left( t \right)} \right]{\rm{,\quad if \quad}}t \le 50\\
\frac{{ - 4s - 2}}{{{s^2} + 2s + 4}}\left[ {u\left( t \right){\rm{ + }}f\left( t \right)} \right]{\rm{, \quad if \quad}}t \ge 50
\end{array} \right.{\rm{,}}
\end{equation}
where the control and disturbance signals were defined as follows:
\begin{equation*}
\begin{array}{l}
u\left( t \right) = {\rm{sin}}\left( {2\pi t} \right) + {\rm{cos}}\left( {3t} \right){\rm{,}}\\
f\left( t \right) = 0.25{\rm{sin}}\left( {0.1\pi t} \right) + {\rm{sin}}\left( {4t} \right){\rm{ + 1}}{\rm{.}}
\end{array}
\end{equation*}

The switched system \eqref{eq27} was considered to demonstrate that the algorithm \eqref{eq26} in some cases is capable of tracking the system parameters change, which is the main motivation for the application of the online identification laws.

For comparison purposes, we also implemented the P-LS estimators \eqref{eq5}, \eqref{eq14} + \eqref{eq18}, the gradient-based estimation law \cite{b1, b2, b3} and the one designed in accordance with the basic DREM procedure \cite{b4}:
\renewcommand{\theequation}{\arabic{equation}a}
\begin{equation}\label{eq29a}
\dot {\hat \theta} \left( t \right) =  - \Gamma \varphi \left( t \right)\left( {{\varphi ^{\top}}\left( t \right)\hat \theta \left( t \right) - z\left( t \right)} \right){\rm{,\;}}\hat \theta \left( {{t_0}} \right) = {\hat \theta _0},
\end{equation}
\setcounter{equation}{22}
\renewcommand{\theequation}{\arabic{equation}b}
\begin{equation}\label{eq29b}
\begin{array}{l}
{{\dot {\hat \theta} }_i}\left( t \right) =  - \gamma \Delta \left( t \right)\left( {\Delta \left( t \right){{\hat \theta }_i}\left( t \right) - {{\cal Y}_i}\left( t \right)} \right){\rm{,\;}}\\
{\cal Y}\left( t \right){\rm{:}} = {\rm{adj}}\left\{ {\Phi \left( t \right)} \right\}Y\left( t \right){\rm{,\;}}\Delta \left( t \right){\rm{:}} = {\rm{det}}\left\{ {\Phi \left( t \right)} \right\}{\rm{,}}\\
\dot Y\left( t \right) =  - lY\left( t \right) + \varphi \left( t \right)z\left( t \right){\rm{,\;}}Y\left( {{t_0}} \right) = {0_{2n}}{\rm{,}}\\
\dot \Phi \left( t \right) =  - l\Phi\left( t \right) + \varphi \left( t \right){\varphi ^{\top}}\left( t \right){\rm{,\;}}\Phi \left( {{t_0}} \right) = {0_{2n \times 2n}}{\rm{,}}
\end{array}
\end{equation}
and, in addition to this, the law on the basis of mixing of the signals \eqref{eq21a} was also implemented:
\setcounter{equation}{22}
\renewcommand{\theequation}{\arabic{equation}c}
\begin{equation}\label{eq29c}
\begin{array}{l}
{{\dot {\hat \theta} }_i}\left( t \right) =  - \gamma \Delta \left( t \right)\left( {\Delta \left( t \right){{\hat \theta }_i}\left( t \right) - {{\cal Y}_i}\left( t \right)} \right){\rm{,\;}}\\
{\cal Y}\left( t \right){\rm{:}} = {\rm{adj}}\left\{ {\psi \left( t \right)} \right\}\vartheta \left( t \right){\rm{,\;}}\Delta \left( t \right){\rm{:}} = {\rm{det}}\left\{ {\psi \left( t \right)} \right\}.
\end{array}
\end{equation}

The parameters of filters \eqref{eq2}, \eqref{eq21a}, \eqref{eq21b}, \eqref{eq29b}, instrumental model \eqref{eq18} and estimation laws \eqref{eq5}, \eqref{eq14} + \eqref{eq18}, \eqref{eq26}, \eqref{eq29a}, \eqref{eq29b}, \eqref{eq29c} were picked as follows:
\renewcommand{\theequation}{\arabic{equation}}
\begin{equation*}
\begin{array}{c}
\Lambda \left( s \right) = R\left( {{\theta _{iv}}{\rm{,\;}}s} \right) = {s^2} + 20s + 100,{\rm{\;}} F_{0}=0.01{\rm{,}}\\
Z\left( {{\theta _{iv}}{\rm{,\;}}s} \right) = 20s + 100,{\rm{\;}}T = 5,{\rm{\;}}p = 10,{\rm{ }}\\
\gamma  = \left\{ \begin{array}{l}
{10^{26}}{\rm{,\;for\;}}\left( {17} \right)\\
{10^{20}}{\rm{,\;for\;}}\left( {23{\rm{b}}} \right)\\
{10^{26}}{\rm{,\;for\;}}\left( {23{\rm{c}}} \right)
\end{array} \right.{\rm{,\;}}\begin{array}{*{20}{c}}
{\Gamma  = 100}{\rm{,\;for\;}}\left( {23{\rm{a}}} \right)\\
{{\Gamma _0} = {{10}^6}{\rm{,\;for\;}}\left( 4 \right)}\\
{{\Gamma _0} = {{10}^3}{\rm{,\;for\;}}\left( {10} \right)}
\end{array},{\rm{\;}}l = 0.{\rm{1}}{\rm{.}}
\end{array}
\end{equation*}

Figure 1 depicts the behavior of the regressor $\Delta \left( t \right)$ and disturbance  $\left\| {{\cal W}\left( t \right)} \right\|$ obtained with the help of equations \eqref{eq21a}, \eqref{eq21b}-\eqref{eq24}.

\begin{figure}[!thpb]
\begin{center}
\includegraphics[scale=0.6]{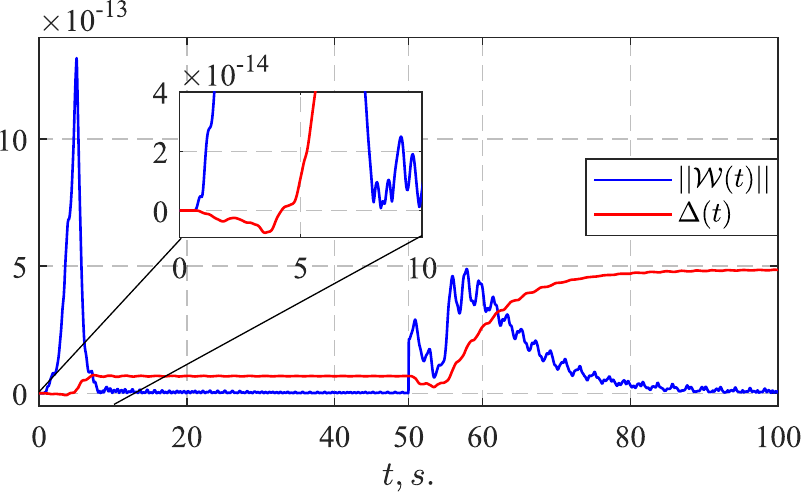}
\caption{{Behavior of $\Delta \left( t \right)$ and $\left\| {{\cal W}\left( t \right)} \right\|$.}} 
\end{center}
\end{figure}

The simulation results presented in Fig. 1 allows one to make the following conclusions:
\begin{enumerate}
    \item [\emph{i})] ${\Delta _{{\rm{UB}}}} \ge \left| {\Delta \left( t \right)} \right|{\rm{\;}}\forall t \ge 0,$
    \item [\emph{ii})] the condition $\left| {\Delta \left( t \right)} \right| \ge {\Delta _{{\rm{LB}}}} > 0$ is met $\forall t \ge 5$,
    \item [\emph{iii})] $\left\| {{\cal W}\left( t \right)} \right\|$ decreases asymptotically, which is required to meet the inequalities $\left| {{{\cal W}_i}\left( t \right)} \right| \le {\textstyle{{\dot F\left( t \right){c_{\cal W}}} \over {F\left( t \right)}}} < \infty .$
\end{enumerate}

Figure 2 presents the comparison of behavior of $\hat \theta \left( t \right)$ obtained by the estimators \eqref{eq26}, \eqref{eq29a}, \eqref{eq29b}, \eqref{eq29c} and \eqref{eq5}, \eqref{eq14} + \eqref{eq18}.

\begin{figure}[!thpb]\label{figure4}
\begin{center}
\includegraphics[scale=0.5]{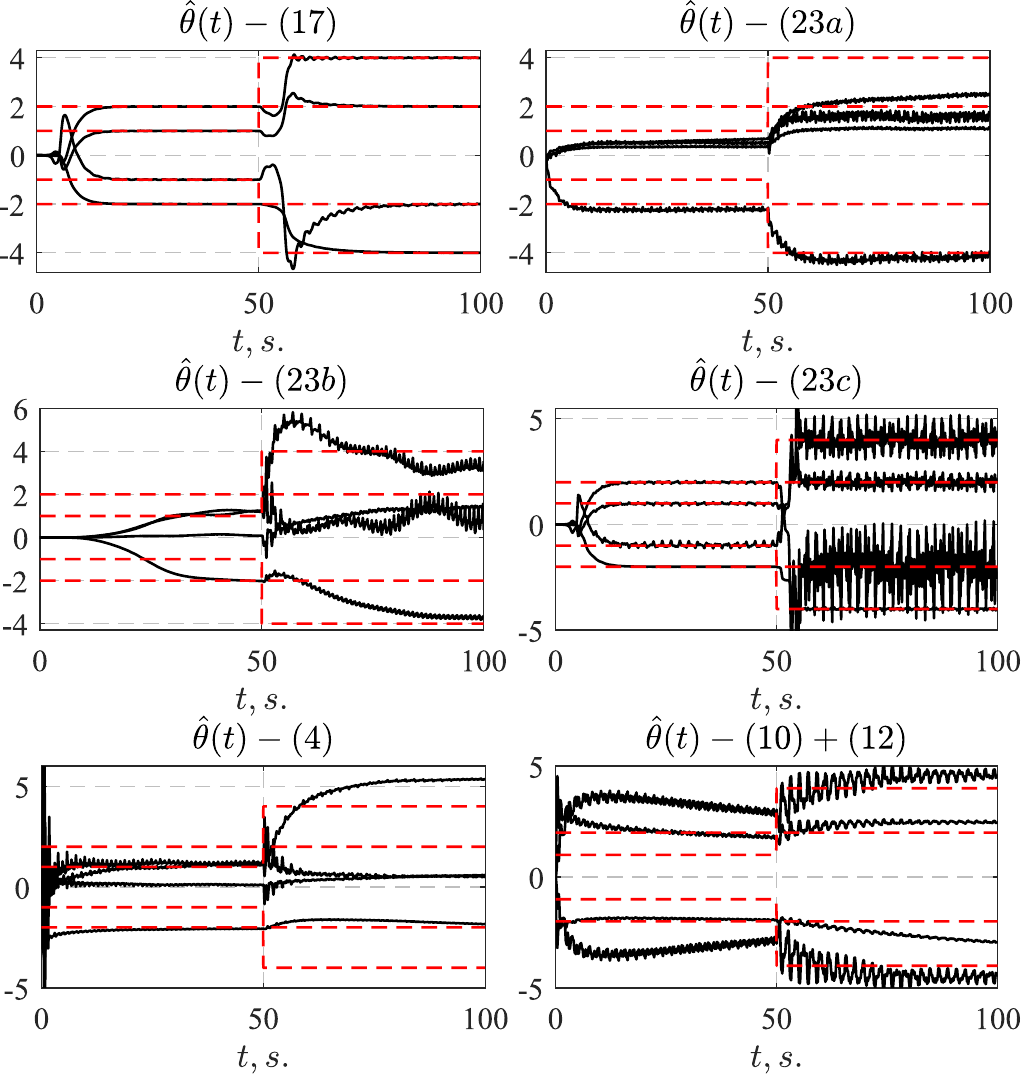}
\caption{{Behavior of  $\hat \theta \left( t \right)$ obtained using estimation laws \eqref{eq26}, \eqref{eq29a}, \eqref{eq29b}, \eqref{eq29c} and \eqref{eq5}, \eqref{eq14} + \eqref{eq18}.}} 
\end{center}
\end{figure}

The simulation results confirm the theoretical conclusions made in the propositions and theorem 1, and show that, considering the laws \eqref{eq26}, \eqref{eq29a}, \eqref{eq29b}, \eqref{eq29c} and \eqref{eq5}, \eqref{eq14} + \eqref{eq18}, only \eqref{eq26} provides exact asymptotic estimation of the unknown parameters $\theta$ in the presence of an unknown external perturbation $f\left( t \right)$. {In comparison with \eqref{eq14} + \eqref{eq18}, the estimates obtained by the proposed law \eqref{eq26} have better transient response and converge to ideal value more rapidly, which is explained by the simpler tuning rules for the proposed law.} The effect and fundamental role of filtering with averaging is clearly seen from comparison of the transients generated by the laws \eqref{eq26} and \eqref{eq29c}.

\section{Conclusion and Further Research}

A new online estimation law for the parameters of linear systems is proposed, which provides the properties \textbf{P1}-\textbf{P3} and ensures asymptotic convergence of the estimation error in the presence of an unknown but bounded perturbation signal, which {meets an independency condition (its spectrum does not have common frequencies with the control/reference signal one).}

The scope of future research is to: \emph{i}) {find a way to increase the rate of convergence and improve alertness to the change of plant unknown parameters,} \emph{ii}) apply the proposed estimation law to the problems of indirect adaptive control and adaptive observers design, \emph{iii}) {find an answer to the question, which was postulated in Section IV.}

\end{document}